# Internal Percolation Problem

I. V. Bezsudnov[1] and A. A. Snarskii[2]

[1] Nauka-Service" Ltd, Moscow, Russia,
[2] National Technical University of Ukraine "KPI", Kyiv, Ukraine

**Abstract.** Introduced and investigated new kind of percolation problem - so called internal percolation problem (IP). In usual percolation current flows from top to bottom of the system and thereby it can be called as external percolation problem (EP) despite of the IP case, when voltage is applied to the system through bars which are inside of the hole in system. EP problem has two major parameters: $M$ – size of system and $a_0$ – size of inclusion, bond size etc. IP holds one more parameter – the size of hole - $L$. Results of computer simulation show that critical indexes of conductance for IP problem are very close to those in EP problem. In opposite, indexes of relative spectral noise density of $1/f$ noise and higher moments are differs from EP problem. The source of such behavior is discussed.

**Keywords:** percolation, 1/f noise, effective conductivity.

**PACS:** 05.60.Cd, 46.65.+g, 64.60.Ak.

## Introduction

Recently, there has been considerable interest in transport processes in randomly disordered media [1-5]. Every 100 days only in Arxiv electronic base there are more than 25 new papers on this subject. Interest in macroscopically disordered media, on the one hand, is due to a large significance of such media - composite materials, and, on the other hand, methods of percolation theory that deal with them are similar to methods of second order phase transitions theory. They describe a great variety of physical processes – from hopping conductivity [3] to pinning of Abrikosov vortices in high-temperature superconductors [5] or superlocalization [6].

The simplest, clean of all extra physical layers, formulation of percolation theory problem is as follows. Given is a lattice of bonds. Part of bonds ($p$) is "black", and remaining part ($1-p$) is "white" (broken bonds). Bonds are colored randomly. To find such a minimum concentration of $p$ "black" bonds, whereby there is a way along the "black" bonds through entire lattice from one "infinity" to another. The same problem can be formulated for lattice sites as well.

Such threshold concentration $p_c$ is called percolation threshold [7]. The existence of percolation threshold is the main "label" or "business card" of percolation theory. The system properties are completely changed when passing through percolation threshold.

In "electric language", "black" bonds (sites) conduct current, "white" (broken) bonds do not (for a more complicated percolation theory, the "blacks" conducts much better than the "whites").

Up to now it has been supposed that in all versions of "electric" percolation problem the contacts are located outside the system on the opposite edges (see Fig 1.a), that defines boundary conditions for determination of local currents and fields in the media.

But in practice shape of microelectronic devices differs from simple rectangle or cube and contacts can be placed on any edge of such structures. To find behavior of effective properties for macroscopically inhomogeneous media with such geometrically non-standard shape and contacts placement is to investigate a new nonstandard class of percolation problems.

On Fig. 1 one can find some possible shape examples of microelectronic devices. Such shapes are now often used in new types of transistors and optoelectronic elements, thick film resistors etc. Will be any difference in behavior of effective properties for nonstandard shape macroscopically inhomogeneous systems? If yes, how shape influence on that?

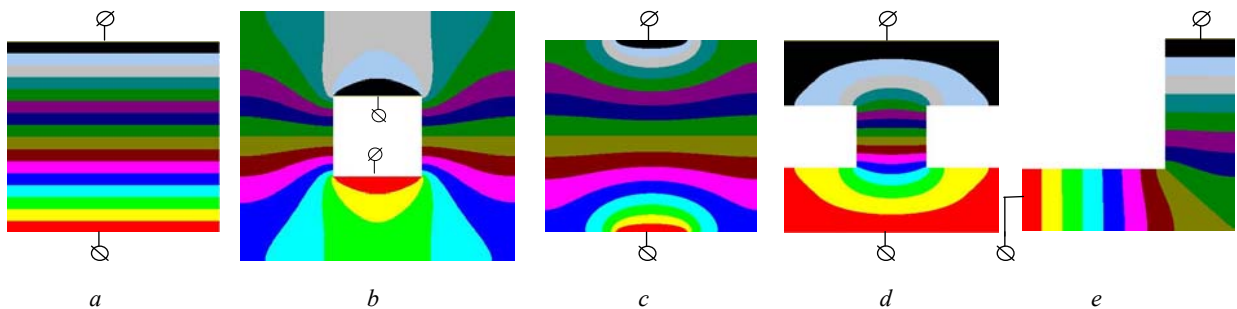

*a*      *b*      *c*      *d*      *e*

**FIGURE 1**. Different shapes with contacts and equipotentials shown, (a) standard (b-e) nonstandard shape (homogeneous case).

In case when contacts are arranged inside the system like on Fig. 1b. percolation problem for such configuration is natural to be called the internal percolation problem (hereinafter IP problem) unlike standard Fig. 1a which in these terms from now on will be called the external (EP problem).

The internal problem is very much similar to the external one. There will be also a very fast change of conductance close to certain concentration of "black" inclusions $p_c$. Conductance behavior near to percolation threshold will be also determined by universal constants – the so-called critical indexes. However, also essential differences exist. In the external percolation problem there are only two characteristic dimensions: $M$ – system size and $a_0$ - typical minimal size, for example, size of grain inhomogeneity or length of lattice cell etc. In the internal problem another size is added (thus, another degree of freedom) – size of hole - $L$. See Fig. 2.

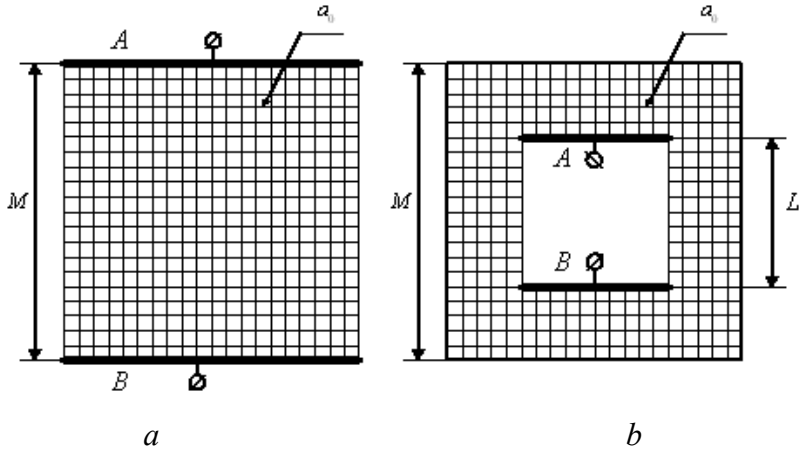

**FIGURE 2**. (*a*) External and (*b*) Internal percolation problems

In EP problem, the condition $M \gg \xi$, where $\xi$ is correlation length [4, 5], means that the value of conductance of the entire media is not depend more on the realization of random structure of media. In IP problem when $M \gg \xi$ two different cases are possible: $L \gg \xi$ and $M - L \gg \xi$, i.e. when size of areas where current flow is more than the correlation length, and realization of random structure has no effect on the system conductance and, on the contrary, when this condition is not met, i.e. either $L \leq \xi$ or $M - L \leq \xi$ and no self-averaging takes place.

One of main parameters of physical processes in inhomogeneous media like conductivity, thermal conductivity, thermoelectricity etc. are so called critical indexes which characterize the behavior of effective kinetic coefficients for such processes. For electric conductivity, when concentration of well conducting phase $p$ is near percolation threshold $p_c$ one can write $\sigma_e(p) \sim (p - p_c)^t$ for $p > p_c$ and $\sigma_e(p) \sim (p_c - p)^{-q}$ for $p < p_c$, where $t$ and $q$ are critical indexes of conductance above and below percolation threshold. Similar relations exist for all of effective kinetic coefficients, for example, Hall coefficient, thermopower etc.

Many numerical experiments (see, for instance, Refs. [3,4,8]) confirm that $t$ and $q$ are independent on lattice type which was used for calculations, i.e. for triangle, square or honeycomb lattice $t$ and $q$ have same values for same dimensions. There was done many attempts [3,4,8] to express $t$, $q$ and other critical indexes of effective kinetic coefficients using only indexes describing geometry of percolation media like critical index of correlation length, size of infinite cluster, circulation radius etc. Many different approximations were developed for critical indexes of kinetic coefficients, that agree more or less with those from numerical or experimental results. But now, generally accepted point of view is that there is no exact analytical link between critical indexes of kinetic coefficients and geometric ones. Moreover it was shown (see, for instance, Refs. [9,10,11], and references therein) that infinite number of critical indexes exists for moments of current distribution on random resistor lattices near percolation threshold.

The paper gives results of numerical simulation of internal percolation problem in 2D and 3D cases. Concentration dependences close to percolation threshold and critical conductivity indexes of IP task and higher moments, in particular, relative spectral density of $1/f$ noise are obtained. It will be shown also that in some cases critical indexes of EP task are differs from IP task and therefore authors state that there exist new universality class of percolation systems.

## 1. IP Problem for Homogeneous Media

First consider a continuous media with specific conductivity $\sigma$ and with minimal size $a_0$ which is negligibly small. In this case if ratio $L/M$ will decrease then the conductance of whole system $G$ also grows, and below certain value $(L/M)_c$ the growth will stop and conductance $G$ will reach the saturation limit $G_c$. Schematically it is shown in Fig. 3a.

The saturation of $G^{IP}$ at $L/M < (L/M)_c$ is based on the fact that major current flows through near-to-hole part of IP system and additional far-from-hole part of media takes less and less current to conduct when $M$ is growing.

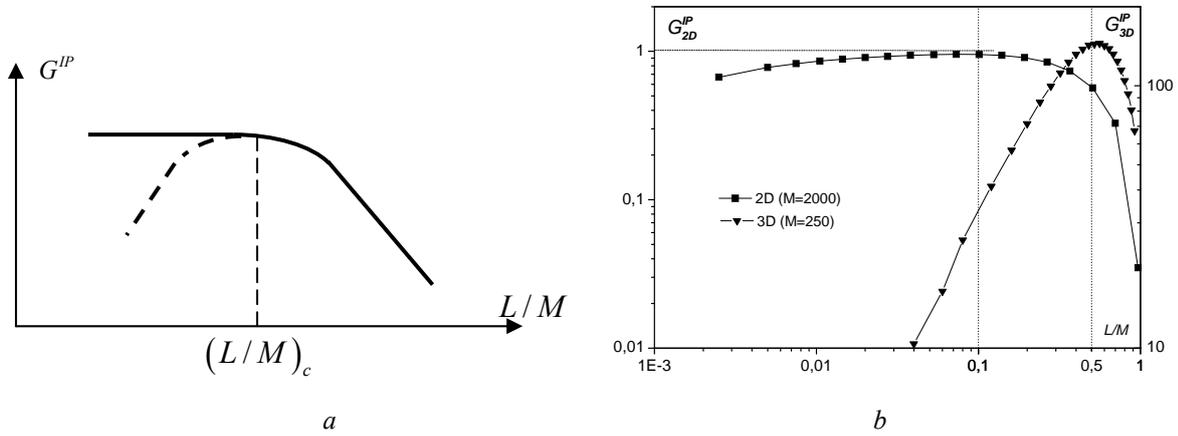

**FIGURE 3.** Conductance of IP system as a function of $L/M$. (*a*) Schematically: Solid line — for continuous media, dashed line — for lattice media. (*b*) Conductance of IP system $G^{IP}$ as a function of $L/M$ in 2D and 3D.

Thus, at $L/M < (L/M)_c$ the external size of the system practically has no influence on the value of full current that flows through the system and, hence, on its conductance. We can say that $M$ becomes "infinite".

In case of non-continuous lattice media at $L/M < (L/M)_c$ conductivity will be determined only by several lattice cells close to current contacts (irrespective of size $M$ at low $L$), and the value of conductance $G$ goes down when $L$ falls (Fig.3a, dashed line).

In EP problem (Fig.2a) the transformation from conductance to specific conductivity value is $\sigma = GM^{d-2}$, where $d$ ($d = 2, 3$) is media dimension, in IP problem such relation is more complicated because it includes the function that takes into account the shape of the hole and the entire media. In this the case (Fig. 2b) this relation can be written in form

$$\sigma = GL^{d-2} f(L, M), \qquad (1)$$

where $f(L,M)$ is dimensionless function.

Determined in Eq. (1) $\sigma$ is just the usual specific conductivity of homogeneous medium. Such form of equation (1) takes into account all geometrical parameters of system, preserves units of all terms of equation and moreover its form is similar to common definition of specific conductivity for EP.

The results of numerical simulation of IP problem for the homogeneous lattice media are given in Fig. 3b, the conductivity of each bond was assumed equal to unity. Dependences $G$ upon ratio $L/M$ for 2D and 3D systems were shown. Calculation proves previously disclosed qualitative pattern of conductance behavior for IP system as a function of its sizes (Fig. 3a).

Conductance $G^{IP}(L/M)$ depend upon $L/M$ ratio, for different values of $M$ $G^{IP}(L/M)$ is a set of equal-shaped curves close to each other, and the less is $M$, the lower curve is. With a two-fold change of $M$ as compared to the values shown in Fig. 3b (for 2D and 3D), the value of conductance $G^{IP}$ at maximum points of the curves is changed no more than by one-two percent, i.e. one can say that with such size of IP system $G$ is no longer depends on $M$.

From Fig. 3b it can be seen that with a choice of $L/M_{2D} = 0.10$ and $L/M_{3D} = 0.5$ the conductance of IP system reaches maximum. With these $L/M$ ratios, conductance $G$ of system with hole size $L$ and $M = \infty$ is almost equal to conductance of system with the same hole and finite external size $M$. Later on calculations were performed precisely with those $L/M$ ratios.

For homogeneous case of IP problem in 2D on Fig 1b. field potential is shown.

N-th moment of Joule heat release of system is calculated as $M(n) = \sum_{\alpha} p_{\alpha}^{n} \Big/ \left( \sum_{\alpha} p_{\alpha} \right)^{n}$, where $p_{\alpha}$ - Joule heat release in $\alpha-th$ bond of lattice. On Fig.4 space distribution of second moment of Joule heat release is shown. Distribution of higher moments is similar to second moment.

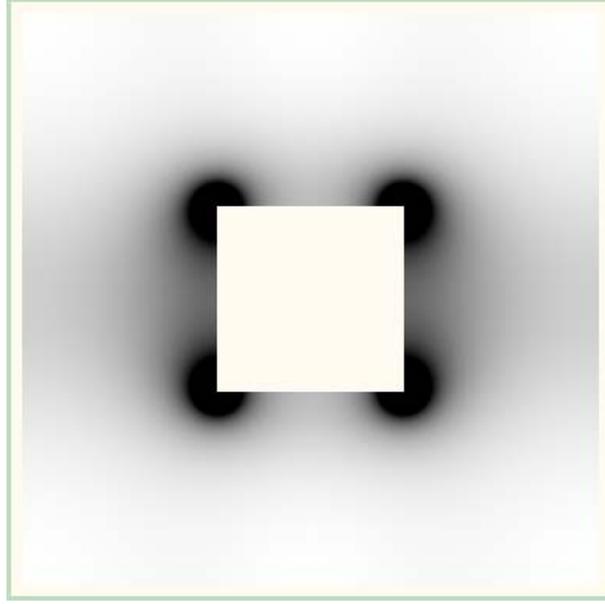

**FIGURE 4**. Space distribution of second Joule heat release moment. Dark places – higher values of local moment.

In case of inhomogeneity calculations were performed in the framework of a standard linear conductivity problem. Numerical experiment was performed on lattices with the total number of sites $\sim 10^6$: 2D lattice with size $M = 1000$ and $L = 100$ and 3D lattice with $M = 100$ and $L = 50$.

Each lattice seat, except for some boundary ones, is assigned a potential that obeys the Kirchhoff's laws. In all cases a bond task was simulated, the resistance of well conducting bonds $r_1$ was assumed equal to unity, and the ratio of resistances of well conducting bonds to poorly conducting bonds $h = r_1 / r_2$. On the two opposite lattice edges A and B (thick lines in Fig. 2) the potential values were set equal to 1 and 0 respectively.

The numerical experiment was performed as follows. The system was produced with uniform randomly distributed well conducting "black" bonds (resistance $r_1$) with concentration $p$ and remaining $(1-p)$ "white" poorly conducting with resistance $r_2$. Then, under the above boundary conditions, a system of Kirchhoff equations was solved and potentials in each lattice seat were found and used to calculate the conductance and the moments of Joule heat release for this realization. Later on they were averaged over 500 realizations in 2D and 100 realizations in 3D.

## 2. Conductance and Effective Conductivity, Percolation Threshold of IP System

In randomly inhomogeneous EP media at $L \gg \xi$ effective specific conductivity $\sigma_e$ behaves the same way as in homogeneous media. Similarly, the effective conductivity of IP system $\sigma_e$ should relate to conductance as in (1) when sizes of system are enough big

$$\sigma_e(p) = G_e^{IP}(p) L^{d-2} f(L, M), \quad L, M \square \ \xi. \tag{2}$$

Concentration dependence of $\sigma_e(p)$ is determined only by $G_e^{IP}(p)$ and, thus in the determination of critical conductivity indexes the geometrical form-factor $L^{d-2} f(L/M)$ is of no importance.

In 2D case the percolation threshold of IP problem corresponds to that of EP problem. This result follows from the fact that with percolation from top to bottom (Fig. 2b), percolation from right to left is impossible, and vice versa. In 3D case percolation threshold was obtained in numerical experiment, and its value coincided to a good accuracy with standard value, for example, from [8].

As for homogeneous case Fig. 4 in inhomogeneous case Fig. 5 local Joule heat release moments are distributed near to hole edges – uneven spatial distribution.

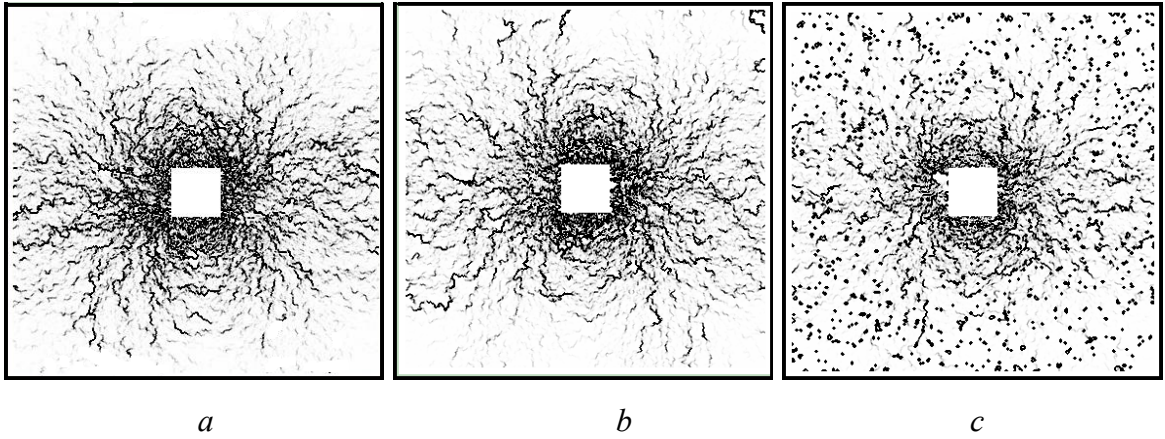

       *a*             *b*             *c*

**FIGURE. 5** Spatial distribution of second moment of Joule heat release in IP problem for $p = 0.55$ and different $h = r_1/r_2$ values (*a*) $h = 10^{-3}$, (*b*) $h = 10^{-5}$, (*c*) $h = 10^{-7}$. Black areas correspond to high values of moment.

### 3. Critical Conductivity Indexes of IP Problem.

Numerical simulation of concentration dependence of conductance $G_e^{IP}$ of IP lattice system was performed for the case of infinitely large inhomogeneity ($h = \infty$) at $p > p_c$. Fig. 6 shows the plots of $G_e^{IP}$ versus $\tau = (p - p_c)/p_c$ - proximity to percolation threshold in log-log scale in 2D and 3D case.

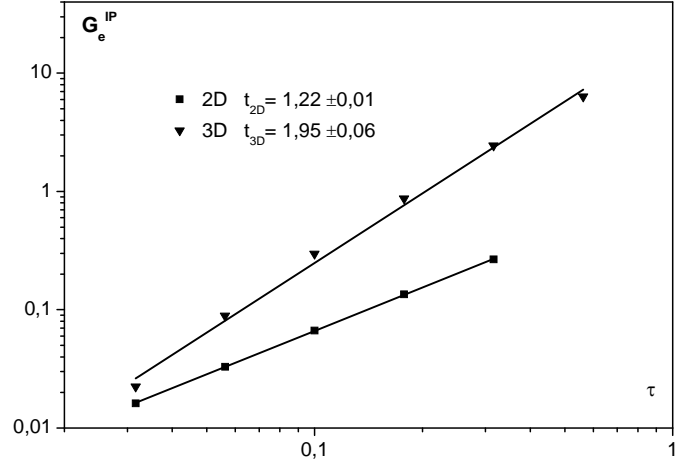

**FIGURE 6.** $G_e^{IP}$ as a function of $\tau$ for 2D and 3D case. The plots also indicate the values of calculated critical conductivity indexes.

Like in EP problem, $G_e^{IP}$ has power behavior dependence on the proximity to percolation threshold and at $p > p_c$ $\sigma_e^{IP} \Box \tau^t$, where $t$ is critical conductivity index.

Using the data obtained in the numerical experiment, conductivity critical indexes of IP problem were calculated in the two and three-dimensions $t_{2D}^{IP} = 1.22$ and $t_{3D}^{IP} = 1.95$. To a good accuracy these indexes coincided with those of EP problem, indicated in the literature Ref. [4].

Thus, behavior of effective conductivity in IP and EP problems are practically identical and the shape has no effect on its behavior.

## 4. $1/f$ Noise and Moments of Joule Heat Release in IP System

In the numerical experiment, along with the calculation of effective conductivity, $1/f$-noise and higher moments were calculated as well.

Critical behavior of moments was studied in many papers, in particular, Refs. [9,12,13] are devoted to investigations of critical indexes of moments behavior in percolation systems. Numerical simulation to determine critical indexes of the moments in 2D and 3D was performed also, see Refs. [14,15].

Results for noise and higher current moments for lattice systems with arbitrary geometry was developed initially in [16], where general formulae are found and lattices made of equal resistors was considered. For the last case authors using Tellegen's and Cohn's theorems show relationship between lattice resistance $R = \sum_\alpha r_\alpha \left(i_\alpha\right)^2 / I^2$, normalized noise $S_R = \{\delta R \delta R\}/R^2$ and moments of current distribution $S_R = s_\alpha \sum_\alpha (i_\alpha)^4 / \left[\sum_\alpha (i_\alpha)^2\right]^2$, where $\{...\}$ time correlation function $r_\alpha$, $s_\alpha$ resistance and noise of each resistor in lattice, $i_\alpha$ current in bond $\alpha$, $I$ current in lattice.

In general case for lattice with arbitrary geometry and arbitrary selected bond resistors, using Cohn's theorem $\delta R = \sum_\alpha \delta R_\alpha (i_\alpha / I)^2$ and uncorrelated resistor fluctuations one obtains following result

$$S_R = \frac{\{\delta R \delta R\}}{R^2} = \frac{\sum_\alpha \frac{\delta r_\alpha \delta r_\alpha}{r_\alpha^2} r_\alpha^2 i_\alpha^4}{R^2 I^4} = \frac{\sum_\alpha s_\alpha p_\alpha^2}{\left(\sum_\alpha p_\alpha\right)^2}, \quad (3)$$

where $p_\alpha$ - Joule heat release in bond $\alpha$.

Specific normalized $1/f$ noise density $C = S \cdot V$, where $V$ is a volume of medium (or number of bonds/seats for lattice system). As the result Eq. (3) can easily be generalized to higher orders

$$C_e(n) = \frac{\sum_\alpha c_\alpha(n) p_\alpha^n}{\left(\sum_\alpha p_\alpha\right)^n}, \quad (4)$$

where $c_\alpha$ - bond property.

Indeed, in all cases $C_e(1) = 1$ for $n = 1$ and $c_\alpha(1) = 1$ for each bond, $C_e(2) = S_R$ - relative spectral noise density of $1/f$ noise for $n = 2$ and $c_\alpha(2) = s_\alpha$, finally for $n \geq 1$ and $c_\alpha(n) = 1$ one gets moments of Joule heat release in the lattice.

In EP problem, see, for example, Refs. [5] and [17] $1/f$ noise and moments have power behavior dependence on the proximity to percolation threshold $C_e(n) \square |\tau|^{-k(n)}$, where $k(n)$ is critical index of $n$-th moment. For EP problem the values of some critical indexes were calculated and cited, for instance, in Ref. [9]. We calculated critical indexes $k(n)$ in EP problem for $n = 2, 3...8$.

Numerical simulation of moments in IP network problem with infinitely large inhomogeneity was performed for the case $p > p_c$. As an example, Fig. 7a shows the plots of the second moment $C_e(2)$ versus $\tau = (p - p_c)/p_c$ in 2D and 3D case. Also on Fig. 4b we show $C_e(2)$ vs $\sigma_e$ of the system. This kind of plots is usually found in experiments.

Main reason to measure experimentally dependence of current moments and particularly $1/f$ noise upon resistance or conductance of whole sample is that the concentration $p$ of phases in measured sample is unknown (exact or even approximately). At the same time moment and conductance of the sample are well measured. In this case current moment and conductance are parametrically defined functions where parameter is the closeness to percolation threshold.

Numerical simulation of such dependencies can be found, for instance, in Refs. [9,13,14,15] and in review [17]. For experimental data see, for instance, Refs. [18-25].

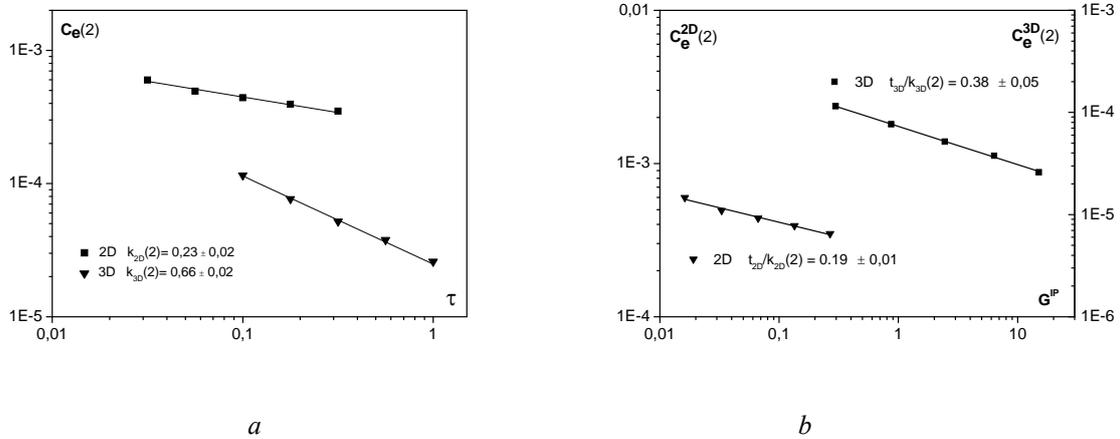

*a*          *b*

**FIGURE 7.** The second moment $C_e(2)$ as a function of (*a*) the proximity to percolation threshold $\tau$ and (*b*) conductance of the system $\sigma_e$.

The table below presents numerical values of conductivity critical index $t$ and indexes $k(n)$ of the moments $C_e(n)$ with $n = 2, 3 \ldots 8$ for EP and IP problem in 2D and 3D cases (precisions for EP problem are not shown).

**TABLE 1.** Numerical values of critical indexes of IP and EP systems.

|  | 2D | | 3D | |
| --- | --- | --- | --- | --- |
|  | IP | EP | IP | EP |
| $t$ | $1.22 \pm 0.01$ | 1.25 | $1.95 \pm 0.06$ | 2.00 |
| $\omega$ | $0.50 \pm 0.005$ | 0.50 | $0.70 \pm 0.005$ | 0.73 |
| $k(2)$ | $0.23 \pm 0.02$ | 1.07 | $0.66 \pm 0.02$ | 1.42 |
| $k(3)$ | -- | 2.35 | $0.95 \pm 0.04$ | 2.94 |
| $k(4)$ | -- | 3.74 | $1.36 \pm 0.07$ | 4.76 |
| $k(5)$ | -- | 5.16 | $1.84 \pm 0.15$ | 6.15 |
| $k(6)$ | -- | 6.65 | $2.34 \pm 0.20$ | 7.57 |
| $k(7)$ | -- | 7.91 | $2.82 \pm 0.22$ | 9.25 |
| $k(8)$ | -- | 9.20 | $3.32 \pm 0.35$ | 10.80 |
|  |  | 1,37 | $0.46 \pm 0.02$ | 1.56 |

Critical indexes of 2D IP problem $k(3) - k(8)$ was not found because of limit of calculation time that are needed to get enough realizations of system to make proper average value.

Below on Fig. 8 are graphically represented the data from Table 1 for IP and EP problems in 3D case.

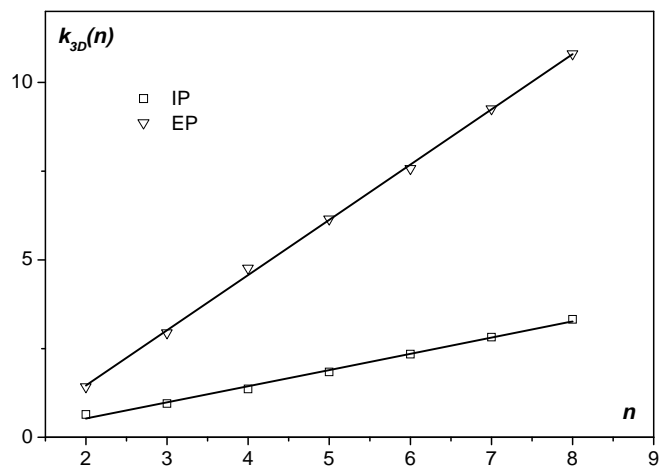

**FIG 8.** The value of critical index $k(n)$ of n-th moment $C_e(n)$ as a function of its number.

It is evident from Fig.8 for EP problem, as it should be Ref. [5], $k(n)$ are linearly dependent on $n$. Note also that critical indexes $k(n)$ in IP problem shows same behavior and they are less than the respective indexes of EP problems in 3D case. Slopes of graphs from Fig.8 calculated and shown on the last line of Table 1.

## 5. IP Problem with Finite Ratio of Conductivities on Percolation Threshold.

As it is known from percolation theory [3] on percolation threshold, in the so-called smearing region, the effective conductivity of media $\sigma_e$ in EP problem

$$\sigma_e = \left(\sigma_1^t \sigma_2^q\right)^{\frac{1}{t+q}}, \qquad (5)$$

where $\sigma_1$ and $\sigma_2$ are specific conductivities of media phases.

In particular, it means that conductance of system with a change in resistance of conductive bonds $r_2$ (all other parameters being constant) behaves as $G \propto r_2^{-\omega}$, $\omega = q/(t+q)$.

Figure below shows the results of numerical simulation of conductance behaviour for IP problem in 2D and 3D case on the percolation threshold.

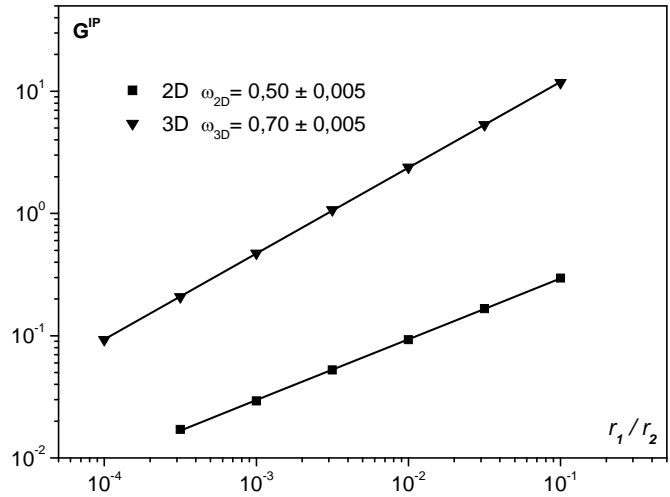

**FIGURE 9.** Conductivity $G^{IP}$ as a function of $r_1/r_2$ at percolation threshold for IP system.

The values of critical indexes calculated for IP problem coincide with the respective critical indexes of EP problem. Just as in the case of metal-insulator system close to percolation threshold, so also in the case of system with finite ratio of conductivities on percolation threshold the behavior of effective conductivity of IP and EP systems is identical.

Table 1 gives critical indexes $\omega$ for EP and IP problems, whence it is seen that they practically coincide. In particular, it means that conductivity critical indexes below percolation threshold - $q$ will coincide with the same accuracy.

## Discussion

As follows from the numerical experiment, critical indexes of EP and IP problems are different from each other, this difference is larger, the higher is the moment number. Critical conductivity indexes are coincides to a good accuracy. We attribute this difference to basically different pattern of the distribution of the Joule heat release in the system or, which is the same, of currents in the system. In EP system there exist both the areas with considerable current concentration and the areas with low current that do not affect the value of the moment being calculated. If it is "difficult" for current to pass through certain area, there can be always another (equivalent) area found which it can "easily" pass. Therefore, for different realizations the location of "difficult" and "easy" places is of little significance.

For IP problem situation is different. The main current flow goes through the areas close to hole edges. Thus, if there are "difficult" places for current flow in this relatively small area, so current will have more problems to find "easy" path because current has to be "removed" from the area close to the hole. The average distribution of currents in the system is basically changed. Moreover the larger is the moment number, the smaller is the area close to the hole where the currents make a major contribution to the moment. Therefore, a change in concentration in IP problem changes not only the number of conductive bonds, "difficult" and "easy" places, like in EP problem, but also the size of the area (close to the hole) wherein the value of moments is "gained". All this leads to change in moments behavior.

Authors believe also that other shape modifications of percolation media, for instance, like on Fig1, can be introduced quite easily and will follow to the rules that are found now for newly introduced IP problem.

## ACKNOWLEDGMENTS

Authors are indebted to M. Zhenirovsky for time consuming discussion of this work and helpful comments.